\documentclass[%
aip,
amsmath,amssymb,
reprint,%
]{revtex4-1}
\usepackage{graphicx}
\usepackage{dcolumn}
\usepackage{bm}
\usepackage[utf8]{inputenc}
\usepackage[T1]{fontenc}
\usepackage{mathptmx}
\usepackage{etoolbox}
\usepackage{ulem}
\usepackage[colorlinks=true,linkcolor=blue]{hyperref}%
\makeatletter
\def\@email#1#2{%
	\endgroup
	\patchcmd{\titleblock@produce}
	{\frontmatter@RRAPformat}
	{\frontmatter@RRAPformat{\produce@RRAP{*#1\href{mailto:#2}{#2}}}\frontmatter@RRAPformat}
	{}{}
}%
\makeatother
\begin{document}
	
	\preprint{AIP/123-QED}
	
	\title{Physics-Guided Interpretable Machine Learning Framework for Anomalous Transport in Crowded Media with Tunable Flexibility}
	\author{Zakiya Shireen}
	\affiliation{Department of Mechanical Engineering, Faculty of Engineering and Information Technology, The University of Melbourne, Victoria,  Australia}
	\author{Sujin B Babu}
	\email{zakiya.shireen@unimelb.edu.au}
	\affiliation{Out of Equilibrium Group, Department of Physics, Indian Institute of Technology Delhi, New Delhi, India}

\begin{abstract}
	Transport in crowded media is governed by the interplay of multiple physical mechanisms. Quantitatively disentangling their individual and coupled contributions remains a longstanding challenge because the evolving microstructure continuously modifies their relative influence. Here, we develop a physics-guided interpretable machine-learning framework that couples Brownian Cluster Dynamics simulations with surrogate machine-learning models and SHAP-based interpretation to quantitatively disentangle the individual and coupled effects of total volume fraction, explorer fraction, and template bond flexibility on explorer-particle transport. We demonstrate the framework using binary colloidal systems in which explorer particles diffuse through a template network formed by irreversible bonds with tunable flexibility. Structural descriptors, mean-squared displacement, intermediate scattering functions, and displacement distributions reveal that network formation localizes explorer particles through enhanced transient caging. Bond flexibility emerges as an independent regulator of post-network relaxation by promoting local bond rearrangements that facilitate the release of transiently caged particles without altering the irreversible network topology. Although crowding and composition dominate the overall transport response, quantitative attribution reveals how the relative contributions of crowding, composition, and template bond flexibility evolve as the confining environment develops. Beyond establishing bond flexibility as a distinct control parameter for relaxation in heterogeneous colloidal networks, this work provides a general physics-guided interpretable machine-learning framework for quantitatively disentangling coupled physical mechanisms in complex transport phenomena.
\end{abstract}

	\maketitle

\section{Introduction}

	The transport of colloidal particles through crowded and heterogeneous environments is central to a broad range of problems in soft matter physics, biological transport, and functional materials design. While classical Brownian motion predicts a linear relationship between the mean-squared displacement (MSD) and time, $\langle \Delta r^2(t) \rangle \sim t$, numerous experimental and theoretical studies have shown that transport in structured media frequently deviates from this behavior. Such deviations, collectively referred to as anomalous diffusion, are characterized by a power-law scaling $\langle \Delta r^2(t) \rangle \sim t^{\alpha}$, where the anomalous exponent $\alpha \neq 1$~\cite{metzler2000random,hofling2013anomalous}. In particular, subdiffusion ($\alpha<1$) commonly arises from crowding, viscoelasticity, and topological constraints that transiently restrict particle motion. These transport characteristics are observed across diverse systems, including the cytoplasm and nucleoplasm of living cells, polymer gels, porous soft materials, and colloidal suspensions, where crowding, intermittent binding, structural confinement, and dynamic heterogeneity govern particle transport ~\cite{regner2013anomalous,banks2005anomalous,hofling2013anomalous,stylianopoulos2010diffusion,watchorn2022untangling,pinholt2021single,seckler2023machine}. Despite their diverse physical origins, the observed dynamics arise from the coupled effects of crowding, composition, and network structure, making it difficult to disentangle the influence of individual physical parameters.
	
	Among the many systems exhibiting anomalous transport, colloidal networks provide particularly versatile model systems because their structure and interactions can be systematically tuned~\cite{kwon2022dynamics, goswami2024anomalous, bechinger2016active}. Variations in network connectivity, mesh size, particle composition, and bond rigidity generate confinement landscapes that strongly influence particle dynamics, giving rise to transient caging, subdiffusion, dynamical arrest, and hopping between locally accessible regions~\cite{mason1997particle, godec2014collective, xu2021enhanced, parrish2017network}. These transport characteristics can be further modified through changes in interparticle interactions, external fields, particle shape, and binary composition, demonstrating the strong sensitivity of anomalous transport to the underlying network morphology~\cite{sheng2021reconfiguring, skora2020macromolecular, koyano2020diffusion, babu2008influence, babu2008tracer, varma2022enhancement, varma2025dimensional}. The wide range of structural and interaction parameters accessible in these systems makes colloidal networks ideal model systems for investigating how surrounding environment influences anomalous transport. However, because transport arises from the coupled effects of network morphology, crowding, composition, and network flexibility, quantifying the contribution of individual physical parameters across multidimensional parameter spaces remains a significant challenge. 

	Given the multifaceted nature of anomalous transport in heterogeneous media, considerable effort has been devoted to developing theoretical and computational approaches that extend beyond classical diffusion models. Analytical frameworks, including generalized Langevin equations, fractional Brownian motion, and continuous-time random walks, have successfully captured many aspects of anomalous transport, including memory effects, viscoelasticity, and non-stationary dynamics~\cite{metzler2000random,milster2024tracer, munoz2021objective}. At the same time, advances in simulation and data-driven analysis have enabled the classification of transport regimes, identification of dynamic heterogeneity, and inference of underlying physical parameters from particle trajectories~\cite{granik2019single, pinholt2021single, seckler2023machine}. However, while these approaches have substantially improved the characterization of anomalous transport, they provide only limited insight into the relative contributions of multiple coupled physical variables governing transport across multidimensional parameter spaces. The increasing complexity of heterogeneous environments has motivated complementary data-driven approaches that analyze transport directly from simulated or experimental trajectories without assuming a specific transport model. Machine-learning methods, including diffusional fingerprinting, unsupervised trajectory clustering, and neural network-based classifiers, have been successfully applied to classify transport regimes, identify dynamic heterogeneity, and infer underlying physical parameters~\cite{cai2025machine, granik2019single, feng2024reliable,shireen2026uncertainty}. However, because many of these approaches emphasize predictive performance, relating their predictions to the underlying physical mechanisms remains challenging. Interpretable machine-learning methods provide a means of addressing this limitation by quantifying the contribution of individual physical parameters to the predicted transport response. Among these, SHapley Additive exPlanations (SHAP) has emerged as a general framework for attributing predictions to individual input features based on cooperative game theory~\cite{lundberg2017unified, lundberg2018consistent}. Recent applications to polymer systems, morphology prediction, and transport in complex fluids have demonstrated its potential for extracting physically interpretable relationships from multidimensional datasets~\cite{shireen2026interpretable,dalal2024polymer,bishnu2025morphology,lellep2022interpreted,shireen2026bayesian}. Despite these advances, most existing applications employ interpretable machine learning primarily as a post hoc analysis tool. A more integrated framework that combines physics-based simulations, surrogate modelling, and quantitative interpretation offers the potential to systematically connect multidimensional parameter spaces with the underlying transport mechanisms.
	
	Here, we develop a physics-guided interpretable machine-learning framework that combines Brownian Cluster Dynamics (BCD) simulations, surrogate machine-learning models, and SHapley Additive exPlanations to investigate anomalous transport in an irreversible binary colloidal system with tunable bond flexibility. Simulations are performed over a multidimensional parameter space spanning total volume fraction, explorer fraction, and bond flexibility, analyzing structural and dynamical observables before and after template-network formation. Rather than relying on machine learning solely for prediction, our integrated framework quantitatively disentangles the individual and coupled effects of these physical parameters, identifies representative physical states, and establishes a quantitative basis for interpreting the resulting transport behavior. By unifying structural descriptors with the mean-squared displacement, intermediate scattering function, and displacement distributions, this approach reveals the fundamental connections between network evolution, transient caging, and explorer relaxation. Beyond the studied binary colloidal system, the proposed framework provides a quantitative methodology for disentangling coupled physical mechanisms across multidimensional transport phenomena.

\section{Methodology}
\begin{figure*}[!ht]
	\centering
	\includegraphics[width=1.0\textwidth]{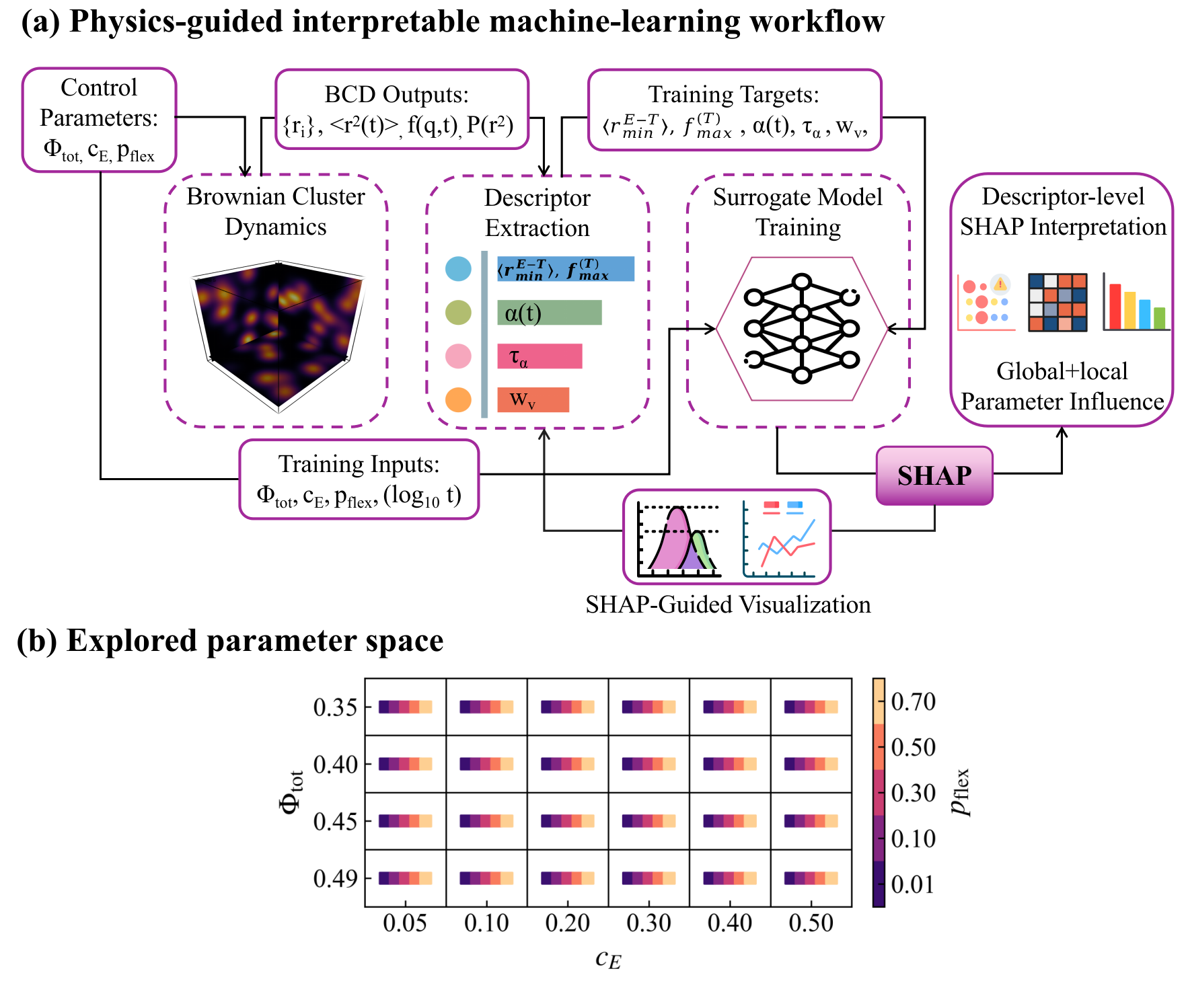}		
	\caption{\textbf{Physics-guided interpretable machine-learning workflow and simulation parameter space.} \textbf{(a)} Schematic illustration of the integrated analysis framework employed in the present study. Brownian Cluster Dynamics simulations are performed over a multidimensional parameter space defined by the total volume fraction ($\Phi_{\mathrm{tot}}$), explorer fraction ($c_E$), and bond flexibility ($p_{\mathrm{flex}}$). The resulting structural configurations and dynamical observables, including representative system snapshots, the mean-squared displacement, intermediate scattering function, and displacement distribution, are used to construct physically meaningful dynamical descriptors such as the anomalous diffusion exponent ($\alpha$), relaxation time ($\tau_{\alpha}$), and displacement-distribution width ($w_v$). These descriptors are subsequently analyzed using surrogate machine-learning models, with SHAP employed to quantify the governing physical descriptors, identify representative dynamical states, and relate multidimensional simulation data to the underlying transport mechanisms. \textbf{(b)} Simulation parameter space explored in the present work, yielding a multidimensional datasets spanning a broad range of confinement and relaxation environments.}
	\label{Figure:1}
\end{figure*}

	\subsection{Explainable Machine-learning framework and SHAP analysis}
	A machine-learning model based on Random Forest regression~\cite{breiman2001random} is constructed to learn the relationship between the simulation control parameters and the structural and dynamical descriptors extracted from the BCD simulations (Fig.~\ref{Figure:1}a). The control parameters comprise the total volume fraction, explorer volume fraction, and bond flexibility ($\Phi_{\mathrm{tot}}$, $c_E$, and $p_{\mathrm{flex}}$), while the target variables comprise the mean minimum explorer--template separation, template-network connectivity quantified by the largest-cluster fraction, mean-squared displacement, local diffusion exponent, relaxation time, and displacement-distribution width ($\langle r_{\mathrm{min}}^{E-T}\rangle$, $f_{\mathrm{max}}^{(T)}$, $\langle r^2(t)\rangle$, $\alpha(t)$, $\tau_{\alpha}$, and $w_v$). For descriptors evaluated over multiple observation times, $\log t$ is included as an additional input variable. This formulation enables the surrogate model to learn the coupled dependence of the extracted descriptors on crowding, composition, bond flexibility and, where appropriate, observation time, without requiring an explicit analytical model. The surrogate models are validated prior to the SHAP analysis to ensure that they provide a reliable basis for subsequent interpretation. The final SHAP analysis is then performed using models trained on the complete datasets. Global SHAP importance, defined as the mean absolute SHAP value over the datasets, ranks the relative influence of the governing physical parameters, while local SHAP values and SHAP dependence evaluations identify state-specific parameter contributions and conditional parameter interactions~\cite{lundberg2017unified,lundberg2018consistent}. Finally, following the SHAP-guided visualization workflow illustrated in Fig.~\ref{Figure:1}a, the SHAP analysis guides the selection of representative simulation states for the structural, MSD, relaxation, and displacement-distribution analyses.
	
	\textbf{Descriptor extraction:}	
	For each simulated state defined by the parameter combination $(\Phi_{\mathrm{tot}}, c_E, p_{\mathrm{flex}})$ and analysis stage, the trajectory-derived observables (MSD, $F(q,t)$, displacement distributions, and morphology snapshots) are converted into a set of structural and dynamical descriptors used for the subsequent sensitivity analysis. Morphology snapshots are analyzed in parallel to quantify the evolving confinement environment through the average minimum explorer-template separation, $\langle r_{\mathrm{min}}^{E-T}\rangle$, and the fraction of template particles belonging to the largest percolated cluster, $f_{\mathrm{max}}^{(T)}$. These descriptors characterize the degree of explorer confinement and the evolution of template-network connectivity before and after network formation. The MSD, averaged over 200 independent realizations, is characterized by its values at selected logarithmically spaced time points together with the corresponding local diffusion exponent, $\alpha(t)$, obtained from the logarithmic slope of the MSD. Together, these descriptors quantify both the extent of particle displacement and the underlying transport regime, with the selected time points capturing the intermediate- and long-time transport dynamics. The intermediate scattering function $F(q,t)$ is characterized by the relaxation time $\tau_{\alpha}$, defined as the time at which $F(q,t)$ decays to $1/e$ of its initial value~\cite{kob1995testing, horbach2001relaxation}. The relaxation time quantifies the characteristic time scale associated with explorer-particle relaxation across the evolving and percolated template network. The displacement distributions are analyzed in the transformed variable $v=\log(r^2)$, from which the distribution width $w_v$ is extracted to characterize the extent of dynamical heterogeneity. Collectively, these descriptors represent the confinement environment, transport dynamics, relaxation behavior, and dynamical heterogeneity of the explorer particles across the investigated parameter space.
	
	\subsection{BCD Simulation Details}	
	Brownian Cluster Dynamics (BCD) simulations were performed using the binary colloid model developed in our previous studies~\cite{shireen2017lattice,shireen2018cage}. The underlying BCD algorithm, including Brownian motion, irreversible bond formation, and cluster construction, follows the original implementation~\cite{meakin1983formation, kolb1983scaling, babu2008influence,shireen2017lattice}. In the present work, the model is modified by incorporating tunable bond flexibility within the template network to investigate its influence on explorer-particle dynamics. The simulated system consists of a binary mixture of hard spheres of diameter $\sigma=1$, comprising freely diffusing explorer particles and network-forming template particles. The explorer concentration is defined as $c_E=N_E/N_{tot}$, while the remaining fraction, $c_T=1-c_E$, forms the template network through irreversible bonding. The composition is varied over $0.05 \leq c_E \leq 0.50$, corresponding to template fractions of $0.95 \geq c_T \geq 0.50$. Throughout the simulations, explorer particles undergo Brownian diffusion, whereas template particles additionally undergo irreversible bond formation to form the template network.
	
	The bond flexibility of the template network is controlled through the parameter $p_{\mathrm{flex}}$, which governs the mobility of bonded template particles while preserving network connectivity. The bond-flexibility protocol follows the implementation described previously~\cite{shireen2023rigidity}. The limiting cases correspond to completely rigid networks ($p_{\mathrm{flex}}=0$), in which bonded template particles remain fixed relative to one another, and fully flexible networks ($p_{\mathrm{flex}}=1$), in which bonded particles undergo Brownian displacement at every movement step while maintaining their bonds. In the present study, the bond flexibility is varied over the range $0.01 \leq p_{\mathrm{flex}} \leq 0.70$ to systematically quantify its influence on explorer-particle dynamics. To distinguish and analyze the influence of network formation from that of the fully developed template network, the simulations are analyzed in two stages. During the \textit{pre-network} stage, explorer-particle dynamics are monitored while the template network evolves through irreversible aggregation. Once network growth reaches a steady state, the final configuration is used to initialize a second simulation, referred to as the \textit{post-network} stage, in which explorer-particle dynamics are analyzed within the fully developed template network.
	
\section{Results and Discussion}
\begin{figure*}[!ht]
	\centering
	\includegraphics[width=0.85\textwidth]{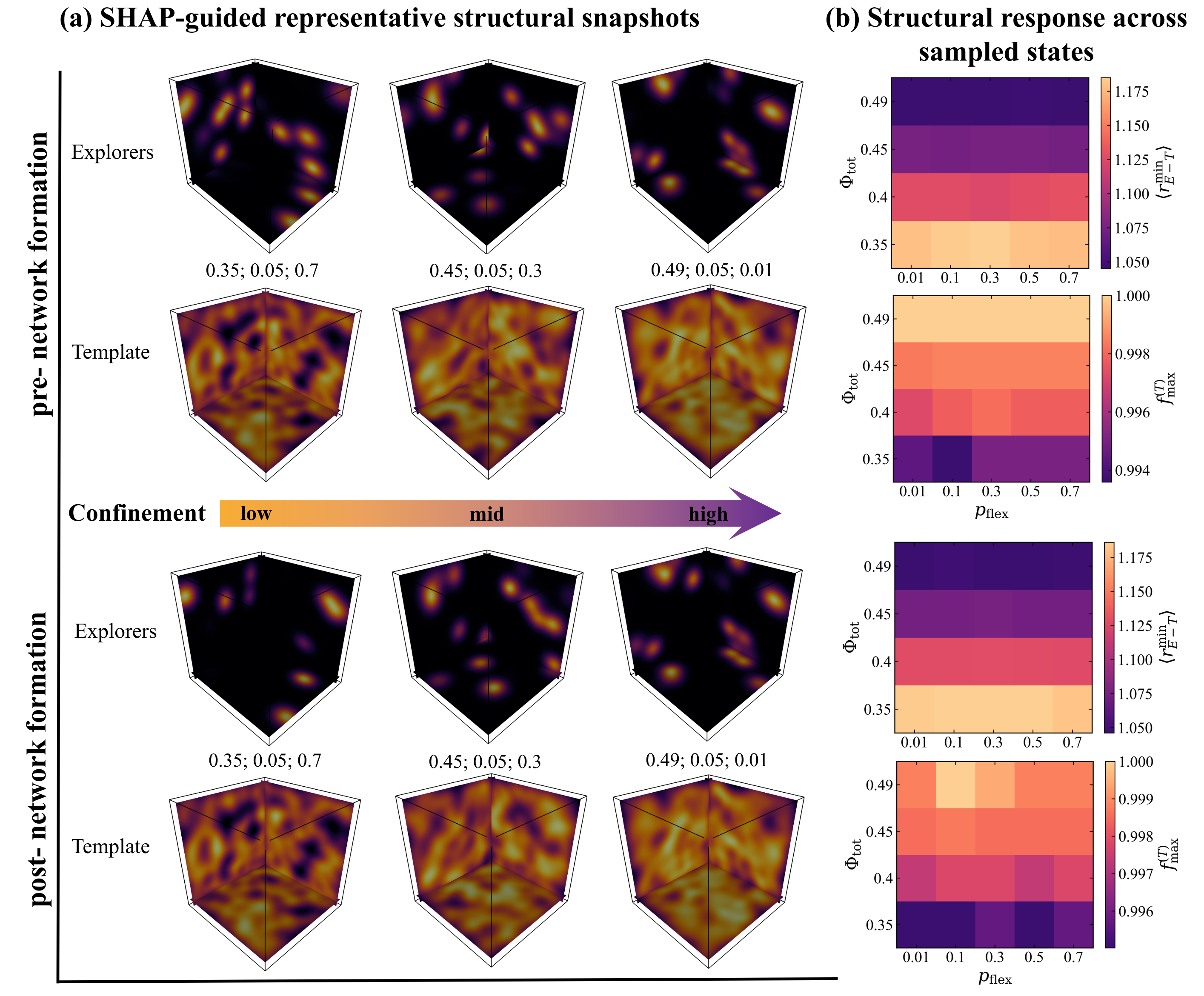}
	\caption{\textbf{SHAP-guided representative structural snapshots and structural response across the sampled parameter space.} \textbf{(a)} Representative explorer-particle and template-network configurations pre- and post- template network formation, selected objectively using the SHAP-guided representative-state analysis described in Fig.~\ref{Figure:1}. For each representative system, the explorer and template snapshots are extracted from the same BCD simulation at the snapshot time ($t\approx10^3$) and rendered as Gaussian-smoothed particle-density maps, where brighter colors indicate regions of higher particle density and darker colours represent low-density regions (free volume). The presented systems span low-, intermediate-, and high-caging environments identified from the SHAP analysis, illustrating the progressive evolution of explorer confinement and template connectivity across the parameter space. The numbers beneath each snapshot denote $(\Phi_{\mathrm{tot}},\,c_E,\,p_{\mathrm{flex}})$. \textbf{(b)} Structural response across the simulated parameter space pre- and post- template network formation. The heat maps show the median values of the two principal structural descriptors used in the SHAP analysis: the average minimum explorer-template separation, $\langle r_{\mathrm{min}}^{E-T}\rangle$, which quantifies the degree of explorer confinement, and the template-network connectivity, quantified by the largest-cluster fraction, $f_{\mathrm{max}}^{(T)}$, plotted as functions of the total volume fraction and bond flexibility.}
	\label{Figure:2}
\end{figure*}	

\subsection{Structural characterization of the confinement environment}	
	To establish the confinement environments experienced by the explorer particles, we first characterize the structure of the binary colloidal system using representative structural snapshots together with quantitative descriptors of explorer confinement and template-network morphology. Figure~\ref{Figure:2} first presents the representative structural configurations and the corresponding structural descriptor maps across the sampled parameter space, followed by a descriptor-level SHAP interpretation of the explorer confinement and template-network descriptors learned from the full parameter space in Fig.~\ref{Figure:3}.
	
\subsubsection{Structural snapshots and descriptor maps}
	Figure~\ref{Figure:2}a presents SHAP-guided visualization of structural snapshots illustrating the binary colloidal system during the pre- and post-network stages. Within each stage, the upper row shows the explorer-particle configuration, while the lower row shows the corresponding template-network configuration from the same simulated system. From left to right, the selected systems span representative low-, intermediate-, and high-confinement environments, reflecting the corresponding changes in template-network morphology and explorer confinement. The representative systems are selected using the SHAP-guided workflow described in Fig.~\ref{Figure:1} and correspond to the same representative confinement environments analyzed throughout the subsequent dynamical sections. Figure~\ref{Figure:2}b complements the snapshot visualizations in Fig.~\ref{Figure:2}a by mapping the corresponding structural descriptors across the sampled states. The heat maps are arranged according to the pre- and post-network stages. Within each stage, the upper heat map corresponds to the explorer-particle row and shows the average minimum explorer-template separation, $\langle r_{\mathrm{min}}^{E-T}\rangle$, while the lower heat map corresponds to the template-particle row and shows the largest-cluster fraction, $f_{\mathrm{max}}^{(T)}$. Both descriptors are plotted as functions of the total volume fraction, $\Phi_{\mathrm{tot}}$, and bond flexibility, $p_{\mathrm{flex}}$, at a fixed explorer fraction of $c_E=0.05$.
	
	The snapshots and corresponding descriptor maps show how the template-network morphology and explorer confinement vary across the pre- and post-network stages. Under low-confinement conditions, the explorer particles remain comparatively well dispersed throughout the accessible volume, while the corresponding template particles form sparsely connected network with large void regions. This behavior is consistent with the larger values of the average minimum explorer--template separation, $\langle r_{\mathrm{min}}^{E-T}\rangle$, and the comparatively smaller values of the largest-cluster fraction, $f_{\mathrm{max}}^{(T)}$, observed at lower $\Phi_{\mathrm{tot}}$. As the confinement increases through the intermediate regime, the template network becomes increasingly interconnected, reducing the accessible free volume available to the explorer particles and causing the local confinement environments~\cite{cohen1959molecular,weeks2000three}. The localization is particularly evident in the intermediate- and high-confinement systems, where the fully formed template network restricts the explorers to smaller accessible regions . For the high-confinement state, the post-network configuration shows stronger localization of the explorers within the developed template structure (see extreme right snapshot for explorers). These changes are accompanied by decreasing $\langle r_{\mathrm{min}}^{E-T}\rangle$ and increasing $f_{\mathrm{max}}^{(T)}$ with $\Phi_{\mathrm{tot}}$, linking the reduced explorer-template separation to the strengthening of template-network connectivity.
	
	These structural changes arise from the coupled influence of the total volume fraction, explorer fraction, and bond flexibility. Increasing $\Phi_{\mathrm{tot}}$ increases the overall particle crowding, promoting the formation of larger connected template clusters while reducing the average separation between explorer and template particles, as reflected by the heatmaps. At a given total volume fraction, decreasing the explorer fraction increases the template concentration, further enhancing network connectivity and restricting the accessible volume available to the explorers. The bond flexibility provides an additional means of controlling the local confinement environment. Lower values of $p_{\mathrm{flex}}$ preserve persistent network structures and maintain stronger confinement, whereas increasing $p_{\mathrm{flex}}$ permits local bond rearrangements that produce more open configurations and larger accessible voids for explorer motion. Together, the structural snapshots and descriptor maps establish how $\Phi_{\mathrm{tot}}$, $c_E$, and $p_{\mathrm{flex}}$ collectively influence the confinement environment experienced by the explorer particles, providing the structural basis for interpreting the corresponding relaxation dynamics and displacement statistics presented in the following sections.

\subsubsection{SHAP interpretation of structural descriptors}
\begin{figure*}[!ht]
	\centering
	\includegraphics[width=0.85\textwidth]{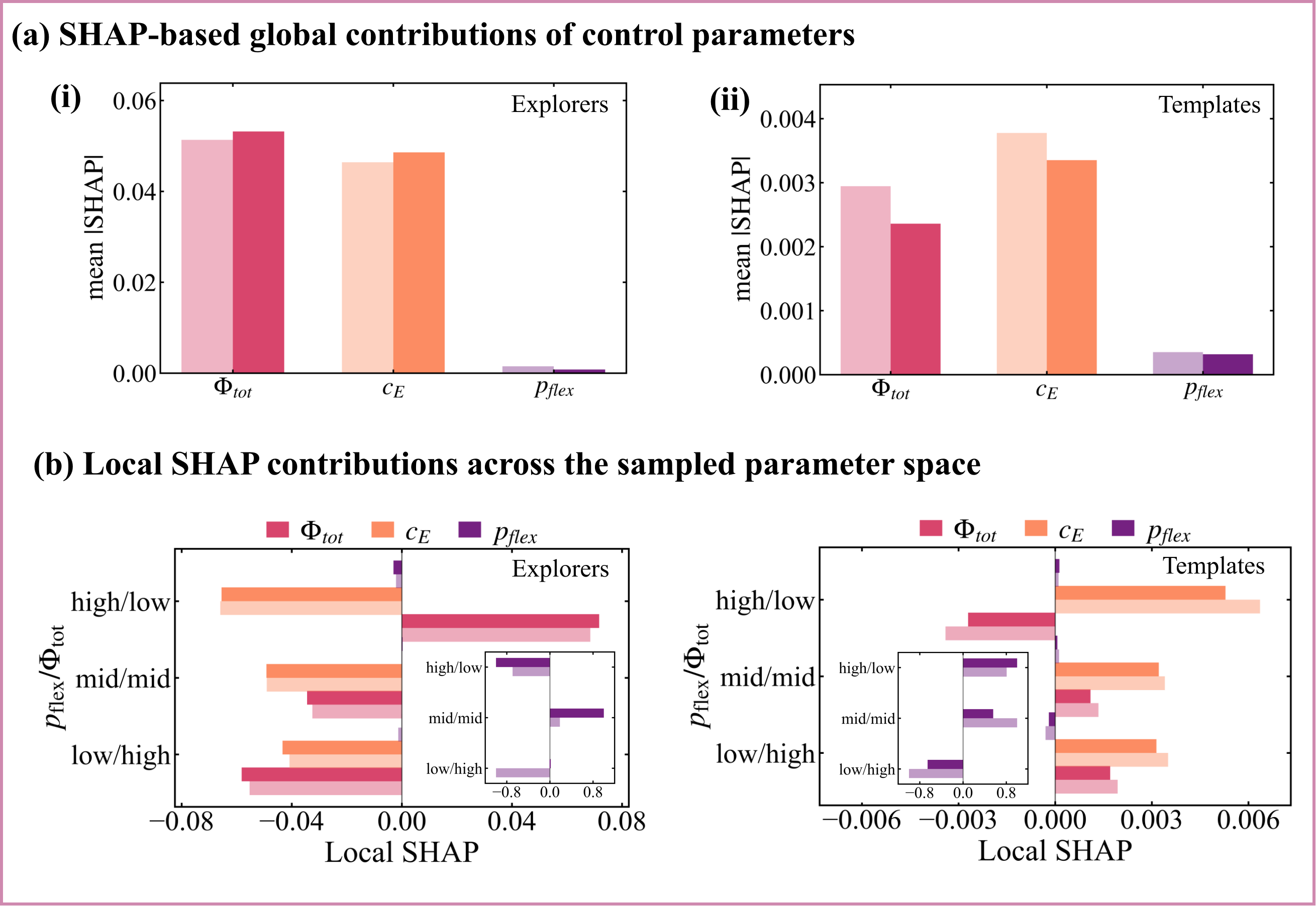}
	\caption{\textbf{SHAP analysis of the structural descriptors governing explorer confinement and template-network formation.}
		\textbf{(a)} Global SHAP importance analysis for the two structural descriptors considered in the present study: (i) the mean minimum explorer-template separation, $\langle r_{\mathrm{min}}^{E-T}\rangle$, and (ii) the fraction of template particles belonging to the largest connected cluster, $f_{\mathrm{max}}^{(T)}$. The bars represent the mean absolute SHAP values associated with the control parameters $\Phi_{\mathrm{tot}}$, $c_E$, and $p_{\mathrm{flex}}$, thereby quantifying their overall contribution to the corresponding structural response across the full parameter space. Light and dark shades correspond to the pre- and post-network states, respectively. \textbf{(b)} Local SHAP contributions for the same structural descriptors evaluated for three representative systems selected from the minimum, median, and maximum SHAP-ranked values of $\langle r_{\mathrm{min}}^{E-T}\rangle$, corresponding to the high/low, mid/mid, and low/high confinement environments, respectively. A normalized inset is shown for the local SHAP contributions of $p_{\mathrm{flex}}$ to improve the visibility of its comparatively smaller contributions.}
	\label{Figure:3}
\end{figure*}

	Figure~\ref{Figure:3}a presents the global SHAP importance analysis of the structural descriptors characterizing explorer confinement and template-network formation. As shown in  Fig.~\ref{Figure:3}a(i), for the explorer confinement metric, $\langle r_{\min}^{E-T} \rangle$, the SHAP ranking identifies the total volume fraction $\Phi_{\mathrm{tot}}$ as the dominant contributor, followed by the explorer fraction $c_E$, with bond flexibility $p_{\mathrm{flex}}$ playing a secondary but non negligible role. This ordering reflects the primary influence of overall crowding and composition on the accessible configuration space of explorers. In contrast, the SHAP importance profile for the template connectivity metric, quantified by the largest cluster fraction $f_T^{\max}$,shown in Fig.~\ref{Figure:3}a(ii), is dominated by composition, with $c_E$ emerging as the most influential parameter, followed by $\Phi_{\mathrm{tot}}$, while $p_{\mathrm{flex}}$ contributes more weakly. For both structural descriptors, the same ordering of the dominant control parameters is retained in pre- and post- network formation, although their relative contributions change modestly. The differing importance profiles indicate that overall crowding dominates explorer confinement, whereas composition exerts the strongest influence on template-network formation, with bond flexibility providing a secondary contribution in both cases.

	Figure~\ref{Figure:3}b presents the local SHAP analysis for three representative confinement environments. These correspond to systems characterized by high bond flexibility and low total volume fraction (high/low), intermediate values of both parameters (mid/mid), and low bond flexibility with high total volume fraction (low/high), spanning the low-, intermediate-, and high-confinement regimes, respectively. The representative systems are selected from the minimum, median, and maximum SHAP-ranked values of the explorer-confinement descriptor, $\langle r_{\min}^{E-T}\rangle$, ensuring that the local SHAP analysis captures physically distinct confinement environments across the explored parameter space. For each system, the local SHAP values quantify how $\Phi_{\mathrm{tot}}$, $c_E$, and $p_{\mathrm{flex}}$ contribute to deviations of the structural descriptors from their dataset-average values. In the explorer-confinement analysis (Fig.~\ref{Figure:3}b(i)), the low-confinement state is dominated by positive contributions from reduced $\Phi_{\mathrm{tot}}$ together with increased $p_{\mathrm{flex}}$, consistent with a more open template network and weaker explorer confinement. The intermediate-confinement state exhibits comparable contributions from crowding and bond flexibility, whereas the high-confinement state is characterized by negative contributions associated with increased $\Phi_{\mathrm{tot}}$ and reduced $p_{\mathrm{flex}}$, reflecting the stronger confinement imposed by a dense interconnected template network. In the corresponding template-network analysis (Fig.~\ref{Figure:3}b(ii)), the local SHAP values for the largest-cluster fraction, $f_{\mathrm{max}}^{(T)}$, show that composition remains the dominant contributor to local variations in template connectivity, followed by $\Phi_{\mathrm{tot}}$, while $p_{\mathrm{flex}}$ contributes more weakly. Additionally, comparison of the pre- and post-network SHAP profiles shows that network formation produces modest changes in the relative contributions of $\Phi_{\mathrm{tot}}$, $c_E$, and $p_{\mathrm{flex}}$ to explorer confinement, whereas the local SHAP profiles governing template-network connectivity remain qualitatively similar across the three confinement environments. Together, the local SHAP analysis shows how the relative contributions of $\Phi_{\mathrm{tot}}$, $c_E$, and $p_{\mathrm{flex}}$ evolve across the three confinement environments, complementing the global trends shown in Fig.~\ref{Figure:3}a.

\subsection{Transport dynamics across the evolving confinement landscape}
\begin{figure*}[!ht]
	\centering
	\includegraphics[width=1.0\textwidth]{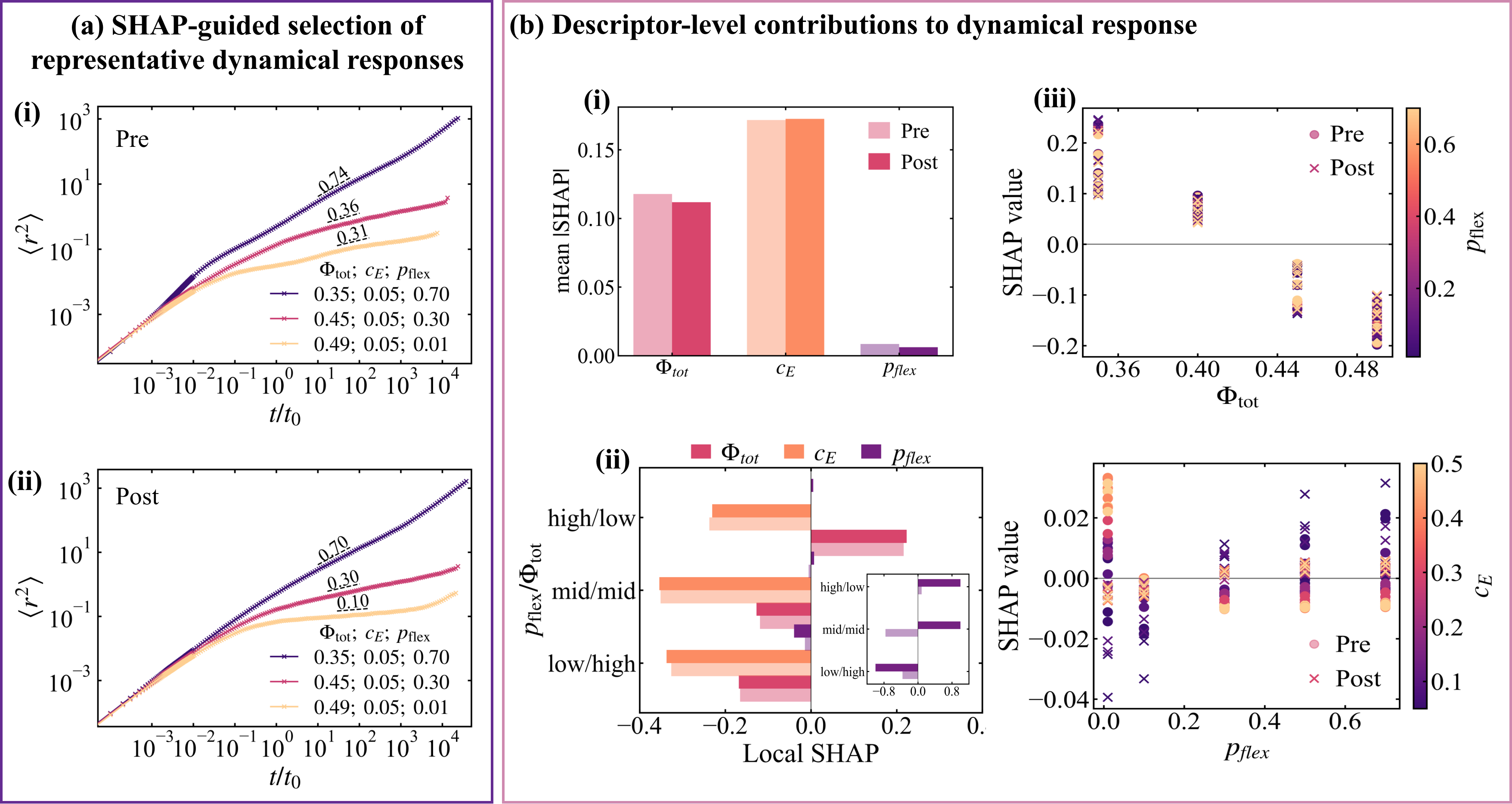}
	\caption{\textbf{Transport dynamics and explainable AI interpretation of explorer diffusion.} (a) SHAP-guided selection of representative explorer mean squared displacement responses illustrating distinct transport regimes before (i) and after (ii) template network formation for systems spanning low-, intermediate-, and high-confinement environments at a fixed explorer fraction ($c_E=0.05$), obtained through different combinations of the total volume fraction ($\Phi_{\mathrm{tot}}$) and bond flexibility ($p_{\mathrm{flex}}$). The corresponding long-time transport exponents ($\alpha$) obtained from the MSD are indicated for each system. (b) SHAP interpretation of the MSD transport exponent, $\alpha(t=10^{3})$, learned from the complete set of simulated binary-colloid systems. Panel (i) shows the global mean absolute SHAP importance of the control parameters before and after template network formation, while panel (ii) presents the corresponding local SHAP explanations for three parameter combinations selected from distinct regions of the learned parameter space. Panels (iii) and (iv) show SHAP dependence plots illustrating how the local contributions of the total volume fraction ($\Phi_{\mathrm{tot}}$) and bond flexibility ($p_{\mathrm{flex}}$), respectively, vary across the parameter space.}
	\label{Figure:4}
\end{figure*}

	To understand how the combined influence of the total volume fraction $\Phi_{\mathrm{tot}}$, explorer fraction $c_E$, and bond flexibility $p_{\mathrm{flex}}$ governs the transport dynamics of binary colloids, we examine the mean squared displacement of explorer particles during (pre-) and after (post-) network formation. Figure~\ref{Figure:4}a presents the mean squared displacement of explorer particles for representative systems selected using the same SHAP-guided sampling strategy employed in the structural analysis. By spanning representative combinations of total volume fraction $\Phi_{\mathrm{tot}}$ and bond flexibility $p_{\mathrm{flex}}$, these systems provide physically interpretable realizations of the structural regimes illustrated in Fig.~\ref{Figure:2}a and quantified through the confinement descriptors in Figs.~\ref{Figure:2}b and~\ref{Figure:3}. All representative systems are shown for a fixed explorer fraction $c_E = 0.05$, while the total volume fraction $\Phi_{\mathrm{tot}}$ and bond flexibility $p_{\mathrm{flex}}$ are varied to probe the coupled influence of crowding and bond flexibility on long-time explorer transport. The upper (~\ref{Figure:4}a(i)) and lower (~\ref{Figure:4}a(ii)) panels correspond to the transport dynamics pre- and post- template network formation, respectively, thereby providing a direct comparison of how the emergence of a persistent template network modifies explorer transport across representative confinement environments.
	
	The MSD curves exhibit a clear separation over several decades in time, reflecting distinct transport regimes associated with the representative confinement environments identified in the structural analysis. Prior to network formation (Fig.~\ref{Figure:4}a(i)), the least confined system ($\Phi_{\mathrm{tot}} = 0.35$, $p_{\mathrm{flex}} = 0.7$) displays the fastest increase in MSD with an effective transport exponent $\alpha \approx 0.74$, indicative of weak confinement and least sub-diffusive motion. Increasing the total volume fraction while reducing bond flexibility progressively suppresses explorer mobility, resulting in sub-diffusive transport with $\alpha \approx 0.36$ for the intermediate confinement state and $\alpha \approx 0.31$ for the most confined system ($\Phi_{\mathrm{tot}} = 0.49$, $p_{\mathrm{flex}} = 0.01$). Following network formation (Fig.~\ref{Figure:4}a(ii)), the separation between the representative transport regimes becomes more pronounced. The least confined system remains the most mobile, with an effective exponent $\alpha \approx 0.70$, whereas the most confined system undergoes a substantial reduction to $\alpha \approx 0.10$. This reduction in the transport exponent after network formation reflects the stronger confinement imposed by the percolated template network, particularly in dense and low-flexibility systems.
	This observation is consistent with confinement-induced transient subdiffusive dynamics reported for tracer particles in flexible gel networks~\cite{godec2014collective}. These representative dynamical responses establish a direct correspondence between the confinement landscapes identified through the structural analysis (Figs.~\ref{Figure:2} and~\ref{Figure:3}) and the resulting explorer transport. While Fig.~\ref{Figure:4}a illustrates representative transport regimes, the SHAP analysis presented in Fig.~\ref{Figure:4}b is performed using the complete set of simulated binary-colloid systems, thereby quantifying how the control parameters govern explorer transport across the full parameter space.
	
	Figure~\ref{Figure:4}b(i) summarizes the global SHAP importance analysis for the MSD transport exponent, $\alpha(t=10^{3})$, performed over the complete set of simulated binary-colloid systems, while Fig.~\ref{Figure:4}b(ii) presents the corresponding local SHAP explanations for three parameter combinations selected from distinct regions of the learned parameter space. Consistent with our previous study of irreversible binary colloids~\cite{shireen2018cage}, the global SHAP analysis identifies the explorer fraction $c_E$ as the dominant parameter governing the MSD transport exponent across both pre- and post-network formation. The total volume fraction $\Phi_{\mathrm{tot}}$ emerges as the second most influential parameter, whereas the contribution of bond flexibility $p_{\mathrm{flex}}$ remains comparatively smaller when averaged over the complete parameter space. This ranking indicates that the overall composition of the binary colloidal system primarily determines the long-time transport response by controlling the relative abundance of mobile explorers and network-forming templates, while the degree of crowding provides a secondary influence through geometric confinement. The local SHAP analysis further demonstrates that the distinct explorer transport observed across the low-, intermediate-, and high-confinement systems does not arise from any single control parameter alone, but from the coupled influence of the total volume fraction $\Phi_{\mathrm{tot}}$, explorer fraction $c_E$, and bond flexibility $p_{\mathrm{flex}}$. Although $c_E$ provides the dominant contribution on average, the balance between the contributions of $\Phi_{\mathrm{tot}}$, $c_E$, and $p_{\mathrm{flex}}$ varies across individual systems, thereby linking the coupled influence of the control parameters to the distinct diffusion behavior illustrated in Fig.~\ref{Figure:4}a.

	Figure~\ref{Figure:4}b(iii) and (iv) further examine how the contributions of the total volume fraction $\Phi_{\mathrm{tot}}$ and bond flexibility $p_{\mathrm{flex}}$ vary across the full parameter space through SHAP dependence analysis. Unlike the global SHAP ranking, which provides parameter importance averaged over all simulated systems, the dependence plots reveal how the influence of a given control parameter changes with the remaining control parameters, thereby exposing coupled effects that are not apparent from the global analysis alone. Specifically, Fig.~\ref{Figure:4}b(iii) examines the contribution of $\Phi_{\mathrm{tot}}$ across different values of bond flexibility $p_{\mathrm{flex}}$, whereas Fig.~\ref{Figure:4}b(iv) shows the corresponding contribution of bond flexibility across different explorer fractions $c_E$. Although the global SHAP ranking identifies $p_{\mathrm{flex}}$ as the least influential control parameter on average, the dependence analysis shows that the local SHAP contributions associated with both $\Phi_{\mathrm{tot}}$ and $p_{\mathrm{flex}}$ vary across the parameter space. Consequently, although $c_E$ remains the dominant global control parameter, the transport response arises from the coupled action of composition, crowding, and bond flexibility. This conditional role of $p_{\mathrm{flex}}$ represents a key distinction from previous studies of irreversible binary colloids, where explorer transport was primarily interpreted in terms of composition and crowding alone~\cite{shireen2018cage}.

\begin{figure*}[!ht]
	\centering
	\includegraphics[width=1.0\textwidth]{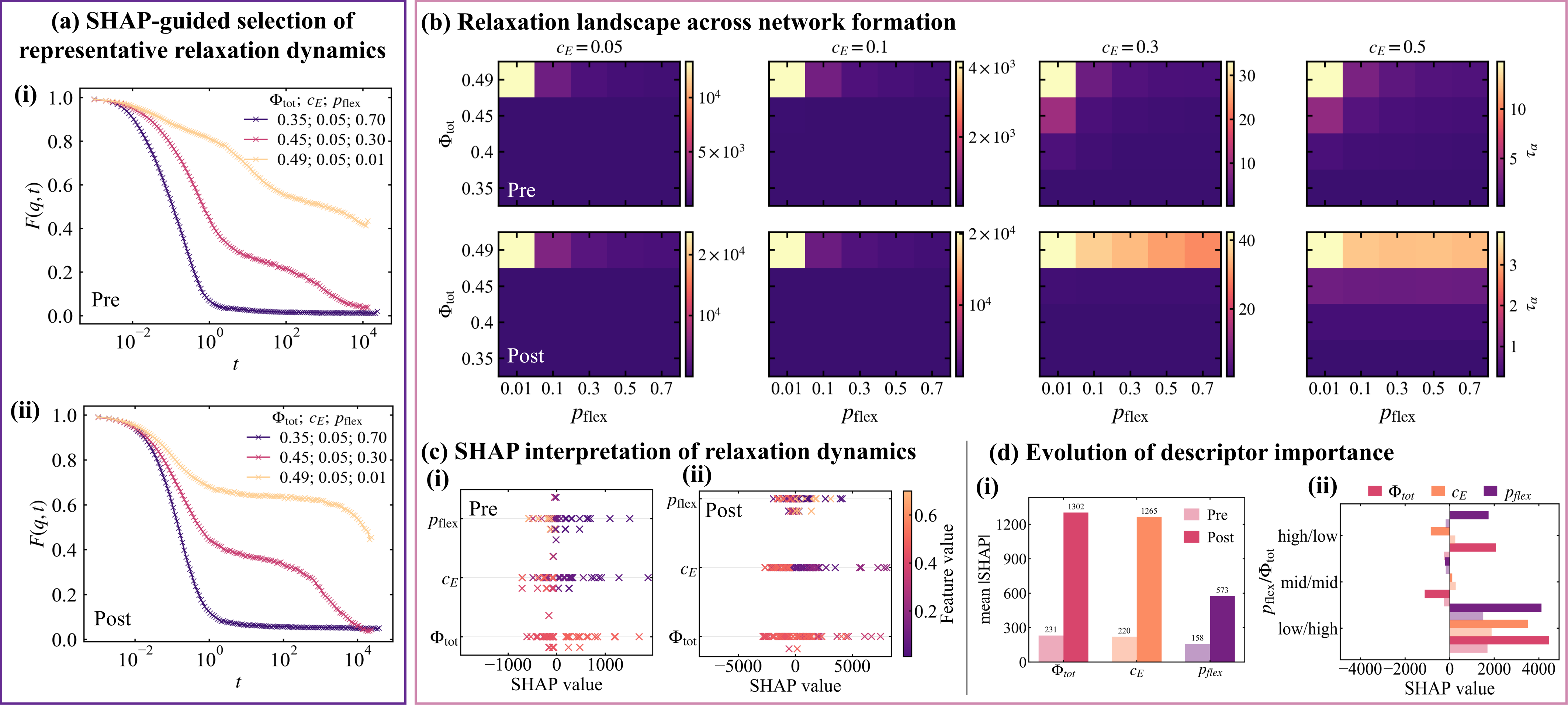}
	\caption{\textbf{SHAP-guided relaxation dynamics and descriptor-level interpretation.} \textit{(a)} SHAP-guided selection of representative intermediate scattering functions, $F(q,t)$, illustrating distinct relaxation dynamics pre- (i) and post- (ii) template network formation for systems spanning low-, intermediate-, and high-confinement environments at a fixed explorer fraction ($c_E=0.05$), obtained through different combinations of the total volume fraction ($\Phi_{\mathrm{tot}}$) and bond flexibility ($p_{\mathrm{flex}}$). \textit{(b)} Relaxation landscape before (top) and after (bottom) template network formation, showing the evolution of the relaxation time, $\tau_{\alpha}$, as a function of the total volume fraction ($\Phi_{\mathrm{tot}}$), bond flexibility ($p_{\mathrm{flex}}$), and explorer fraction ($c_E$). Each column corresponds to a different explorer fraction ($c_E=0.05$, $0.10$, $0.30$, and $0.50$), with the color scale in each heatmap representing the corresponding relaxation time, $\tau_{\alpha}$. \textit{(c)} SHAP interpretation of the relaxation dynamics pre- (i) and post- (ii) network formation. The SHAP summary (beeswarm) plots illustrate the global distribution of feature contributions to the relaxation time, $\tau_{\alpha}$. Within each row, the marker color represents the value of the corresponding descriptor, with darker and lighter shades indicating lower and higher descriptor values, respectively. \textit{(d}) Evolution of descriptor importance pre- and post-template network formation. Panel (i) shows the mean absolute SHAP importance of the control parameters before and after network formation, with the numerical values above the bars indicating the corresponding mean absolute SHAP values. Panel (ii) presents the corresponding local SHAP contributions for the three representative confinement environments identified through the SHAP-guided sampling strategy. In both panels, the lighter and darker shades represent the pre- and post-network formation systems, respectively, whereas the colors distinguish the three control parameters: $\Phi_{\mathrm{tot}}$, $c_E$, and $p_{\mathrm{flex}}$.}
	\label{Figure:5}
\end{figure*}	

\subsection{Relaxation behavior and its descriptor-level interpretation}
	To further examine the dynamical origin of the transport behavior identified from the MSD analysis, we next investigate the relaxation behavior of explorer particles through the intermediate scattering function, $F(q,t)$. Figure~\ref{Figure:5}(a) first compares the relaxation behavior pre- and post- template network formation for systems spanning distinct confinement environments, followed by a descriptor-level SHAP interpretation of the relaxation time, $\tau_{\alpha}$, learned from the full parameter space in Figs.~\ref{Figure:5}(b) and (c). To provide physical insight into the descriptor-level relationships identified by the SHAP analysis, Fig.~\ref{Figure:6} presents additional $F(q,t)$ comparisons that isolate the coupled effects of composition, crowding, and bond flexibility on the relaxation dynamics.
	
\subsubsection{SHAP-guided interpretation of the intermediate scattering function}
	Figure~\ref{Figure:5}a presents the intermediate scattering function, $F(q,t)$, for representative systems spanning low-, intermediate-, and high-confinement environments selected by using SHAP-guided sampling strategy. The upper (Fig.~\ref{Figure:5}a(i)) and lower (Fig.~\ref{Figure:5}a(ii)) panels correspond to the relaxation dynamics pre- and post- network formation, respectively. The self-intermediate scattering function exhibits a systematic slowing with increasing confinement, consistent with the transport behavior identified from the MSD analysis. As shown in Fig.~\ref{Figure:5}a(i), prior to network formation the least confined system ($\Phi_{\mathrm{tot}} = 0.35$, $p_{\mathrm{flex}} = 0.70$) relaxes rapidly through a single-step decay. At intermediate confinement, increasing the total volume fraction and reducing bond flexibility delay the relaxation, leading to the emergence of a plateau and the onset of two-step relaxation ($\beta$ and $\alpha$)~\cite{gotze1992relaxation}. Under high-confinement conditions, the plateau shifts to a higher correlation level and persists over longer times before the system undergoes the final relaxation. The corresponding post-network relaxation dynamics are shown in Fig.~\ref{Figure:5}a(ii), where the separation between the relaxation responses becomes more pronounced. While the least confined system continues to exhibit rapid single-step relaxation, increasing confinement further extends the plateau regime and shifts the final $\alpha$-relaxation to longer times. In the highly confined system, the plateau becomes both higher and longer-lived, indicating that the formation of a dense, percolated network enhances the transient localization of explorer particles before the eventual relaxation. 
	
	The observed evolution of the relaxation dynamics reflects the coupled influence of composition, crowding, and bond flexibility on the local confinement experienced by explorer particles. At low total volume fractions, the larger free volume and higher bond flexibility allow explorer particles to decorrelate rapidly, giving rise to a predominantly single-step relaxation. Increasing the total volume fraction reduces the available free volume, while lower bond flexibility suppresses local network rearrangements, resulting in stronger transient caging and the emergence of a well-defined two-step relaxation. Following template network formation, the percolated template network further restricts the accessible pathways for explorer motion, thereby enhancing particle localization and extending the lifetime of the plateau before the final $\alpha$-relaxation. A similar relationship between network connectivity, rigidity, and collective behavior has recently been reported for space-spanning attractive colloidal gels~\cite{nabizadeh2024network}.
	
	Figure~\ref{Figure:5}b extends the descriptor-level analysis by reconstructing the relaxation landscape learned from the full simulation datasets. The relaxation time, $\tau_{\alpha}$, is mapped over the full parameter space pre- and post- template network formation. Each column corresponds to a fixed explorer fraction, $c_E$, while the relaxation time is shown as a function of the total volume fraction, $\Phi_{\mathrm{tot}}$, and bond flexibility, $p_{\mathrm{flex}}$. The upper and lower rows represent the pre- and post- network formation systems, respectively. Prior to network formation (top row), the relaxation landscape is primarily governed by the total volume fraction. Across all explorer fractions, increasing $\Phi_{\mathrm{tot}}$ systematically increases the relaxation time, indicating slower relaxation with increasing crowding. In contrast, the influence of bond flexibility remains comparatively weak, with variations in $p_{\mathrm{flex}}$ producing only modest changes in $\tau_{\alpha}$ over the entire flexibility range. Consequently, the strongest relaxation slowdown is consistently observed at the highest total volume fraction, whereas the least crowded systems remain rapidly relaxing irrespective of bond flexibility. Following template network formation (bottom row), the relaxation landscape changes qualitatively. Although the total volume fraction continues to dominate the relaxation dynamics, the influence of bond flexibility becomes increasingly pronounced with increasing explorer fraction. At low explorer fractions ($c_E=0.05$ and $0.10$), the relaxation behavior remains largely controlled by crowding, with only a weak dependence on $p_{\mathrm{flex}}$. However, at higher explorer fractions ($c_E=0.30$ and $0.50$), a clear flexibility-dependent gradient emerges, where increasing bond flexibility systematically reduces $\tau_{\alpha}$, particularly at high total volume fractions. This trend indicates that flexible template networks facilitate local structural rearrangements and accelerate relaxation dynamics despite the presence of the percolated network. The evolution of the relaxation landscape therefore shows that the influence of bond flexibility is itself regulated by the system composition. More broadly, these descriptor-level relationships reinforce recent efforts to establish quantitative links between local structure and relaxation dynamics in dense colloidal systems~\cite{sahu2024structural}. Before network formation, the absence of a persistent template network limits the role of bond flexibility in controlling the relaxation dynamics. Once the template network is established, however, bond flexibility becomes an increasingly effective mechanism for relieving local confinement, particularly in explorer-rich systems where frequent explorer-template interactions promote local network rearrangements, thereby facilitating the relaxation of transient cages. Consequently, network formation not only enhances the overall confinement experienced by explorer particles but also introduces a flexibility-dependent relaxation pathway whose influence becomes stronger with increasing explorer fraction.
	
	To understand the physical origin of the machine-learned relaxation landscape shown in Fig.~\ref{Figure:5}b, Fig.~\ref{Figure:5}c presents the corresponding SHAP summary plots before (i) and after (ii) template network formation. The SHAP distributions reveal that the total volume fraction remains the dominant descriptor governing the learned dependence of the relaxation time, with larger values of $\Phi_{\mathrm{tot}}$ predominantly associated with positive SHAP values and lower values with negative SHAP values. This behavior indicates that increasing crowding systematically shifts the relaxation toward longer times, consistent with the relaxation landscapes in Fig.~\ref{Figure:5}b. In contrast, the SHAP contributions associated with the explorer fraction remain more broadly distributed and exhibit a comparatively weaker dependence than those of the total volume fraction. Nevertheless, the distribution is consistent with the dependence of the relaxation time on explorer fraction observed in Fig.~\ref{Figure:5}b. Prior to template network formation, the SHAP distribution associated with bond flexibility remains comparatively narrow, indicating that variations in $p_{\mathrm{flex}}$ exert only a limited influence on the learned relaxation behavior. Following template network formation, however, the SHAP distribution broadens substantially, demonstrating that the relaxation dynamics becomes increasingly sensitive to bond flexibility. This transition is consistent with the relaxation landscapes in Fig.~\ref{Figure:5}b, where the influence of $p_{\mathrm{flex}}$ evolves from being comparatively weak prior to network formation to becoming more pronounced with increasing explorer fraction after the formation of system-spanned percolated template network.

	The evolution of the global and representative local SHAP contributions is summarized in Fig.~\ref{Figure:5}d. The global descriptor importance shown in Fig.~\ref{Figure:5}d(i) confirms that the total volume fraction remains the dominant descriptor governing the relaxation both before and after template network formation. The explorer fraction exhibits the second-largest overall importance, whereas bond flexibility contributes less strongly in the pre-network system but becomes substantially more influential following network formation. This increase in the absolute SHAP contribution of $p_{\mathrm{flex}}$ is fully consistent with the enhanced sensitivity to bond flexibility observed in both the relaxation landscapes (Fig.~\ref{Figure:5}b) and the SHAP summary distributions (Fig.~\ref{Figure:5}c), reflecting the increasing role of the template network mechanics in controlling the relaxation behavior. This trend is further supported by the representative local SHAP contributions in Fig.~\ref{Figure:5}d(ii), which reveal how the relative influence of the descriptors varies across different confinement environments. For the highly flexible, low-crowding system (high/low), the dominant positive contributions arise from the total volume fraction and bond flexibility, whereas the contribution from the explorer fraction remains comparatively weak and changes only modestly following network formation. As the system approaches the intermediate confinement regime (mid/mid), the descriptor contributions become comparatively smaller, indicating that the effects of crowding, composition, and bond flexibility largely compensate one another. In contrast, for the least flexible and most crowded system (low/high), the positive contributions from all three descriptors increase markedly following template network formation, with the total volume fraction remaining the dominant contributor while the contribution of bond flexibility increases substantially relative to the pre-network formation state. Together, these results demonstrate that the relaxation of explorer particles is governed primarily by the degree of crowding, whereas the formation of a percolated template network enhances the role of bond flexibility by regulating the lifetime of transient cages and, consequently, the overall relaxation behavior.
		
\subsubsection{Origin of the flexibility-dependent relaxation landscape}
\begin{figure*}[!ht]
	\centering
	\includegraphics[width=0.75\textwidth]{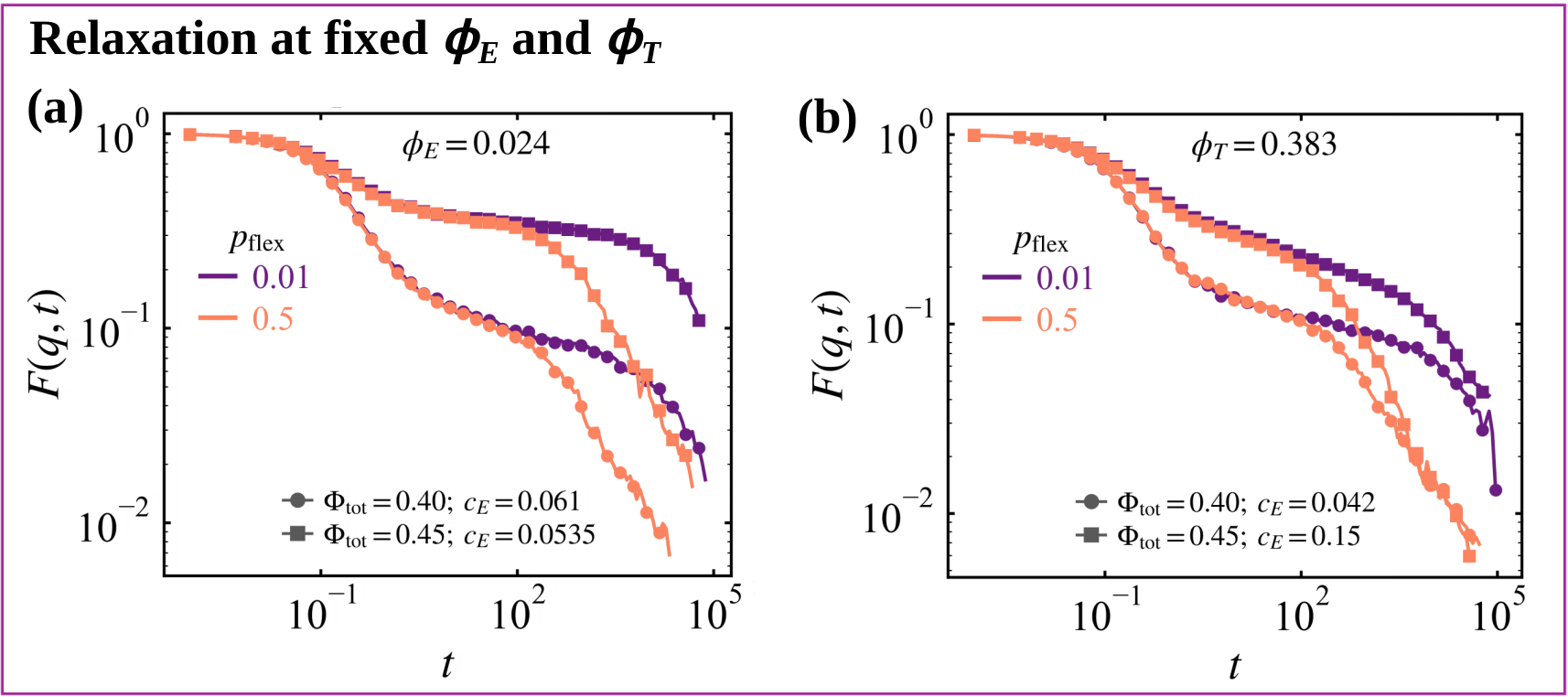}
	\caption{\textbf{Post-network relaxation dynamics at fixed explorer and template volume fractions.} \textbf{(a)} Intermediate scattering functions of explorer particles for representative post-network systems at a fixed explorer volume fraction, $\phi_{\mathrm{E}}=0.024$, illustrating the effect of bond flexibility for two total volume fractions. Holding $\phi_{\mathrm{E}}$ constant isolates the influence of increasing template volume fraction with increasing total crowding. Circular and square symbols correspond to $\Phi_{\mathrm{tot}}=0.40$ and $0.45$, respectively, while purple and orange curves denote rigid ($p_{\mathrm{flex}}=0.01$) and flexible ($p_{\mathrm{flex}}=0.5$) template networks. \textbf{(b)} Corresponding relaxation dynamics at a fixed template volume fraction, $\phi_{\mathrm{T}}=0.383$, for two combinations of total and explorer volume fractions. Maintaining a constant template fraction isolates the influence of composition while preserving the extent of the confining template network. Symbols and colors are defined as in panel (a).}
	\label{Figure:6}
\end{figure*}	

	To elucidate the physical origin of the enhanced bond-flexibility dependence identified by the SHAP analysis following template network formation (Figs.~\ref{Figure:5}b--d), the post-network relaxation dynamics are compared under controlled confinement environments in Fig.~\ref{Figure:6}. To disentangle the effects of composition and crowding, the relaxation dynamics are compared at fixed explorer volume fraction, $\phi_{\mathrm{E}}=0.024$ (Fig.~\ref{Figure:6}a), and fixed template volume fraction, $\phi_{\mathrm{T}}=0.383$ (Fig.~\ref{Figure:6}b), while varying the bond flexibility, $p_{\mathrm{flex}}$. In both cases, increasing the total volume fraction raises the plateau height, reflecting enhanced transient localization arising from stronger caging within the increasingly crowded template network. The influence of bond flexibility, however, emerges predominantly during the second step of the relaxation, where local bond rearrangements facilitate the release of explorer particles from these transient cages. This delayed relaxation reflects the increasing importance of local structural rearrangements that enable explorer particles to escape transient cages, giving rise to the hallmark two-step relaxation of glassy dynamics~\cite{kob1995testing}, and is closely related to recent observations of rearrangement-assisted penetrant transport in dynamically cross-linked polymer networks~\cite{layding2026small}. When the explorer volume fraction is held fixed, increasing the total volume fraction increases the template volume fraction, leading to a increasingly stronger dependence of the long-time relaxation on bond flexibility, as reflected in the distinct long-time relaxation of the rigid and flexible template networks. This demonstrates that the influence of bond flexibility on the long-time relaxation becomes increasingly pronounced as the template network occupies a larger fraction of the system. In contrast, when the template volume fraction is held fixed, the relaxation curves corresponding to the same bond flexibility approach a similar long-time decay despite the different combinations of $\Phi_{\mathrm{tot}}$ and $c_E$. This indicates that, once the extent of the confining template network is fixed, variations in composition alone do not substantially modify the influence of bond flexibility on the explorers' relaxation dynamics. Instead, the relaxation profile is governed by the cooperative interplay between the total volume fraction, which determines the total crowding, the template volume fraction, which defines the extent of the confining network, and the bond flexibility, which regulates local network rearrangements responsible for the release of explorer particles from these transient cages.
	
	The SHAP analysis in Fig.~\ref{Figure:5}b revealed that bond flexibility becomes substantially more important following template network formation, as evidenced by the pronounced flexibility-dependent relaxation landscape (Fig.~\ref{Figure:5}b), the broadening of the SHAP contribution associated with $p_{\mathrm{flex}}$ (Fig.~\ref{Figure:5}c), and the corresponding increase in its global SHAP importance (Fig.~\ref{Figure:5}d). Figure~\ref{Figure:6} provides the mechanistic basis for these machine-learned relationships by demonstrating how crowding, composition, and bond flexibility cooperatively govern the post-network relaxation dynamics. The descriptor-level relationships identified by the SHAP model are therefore consistent with the underlying relaxation physics revealed by the controlled dynamical comparisons.

\subsection{Displacement distribution signatures of relaxation dynamics}
\begin{figure*}[!ht]
	\centering
	\includegraphics[width=1.0\textwidth]{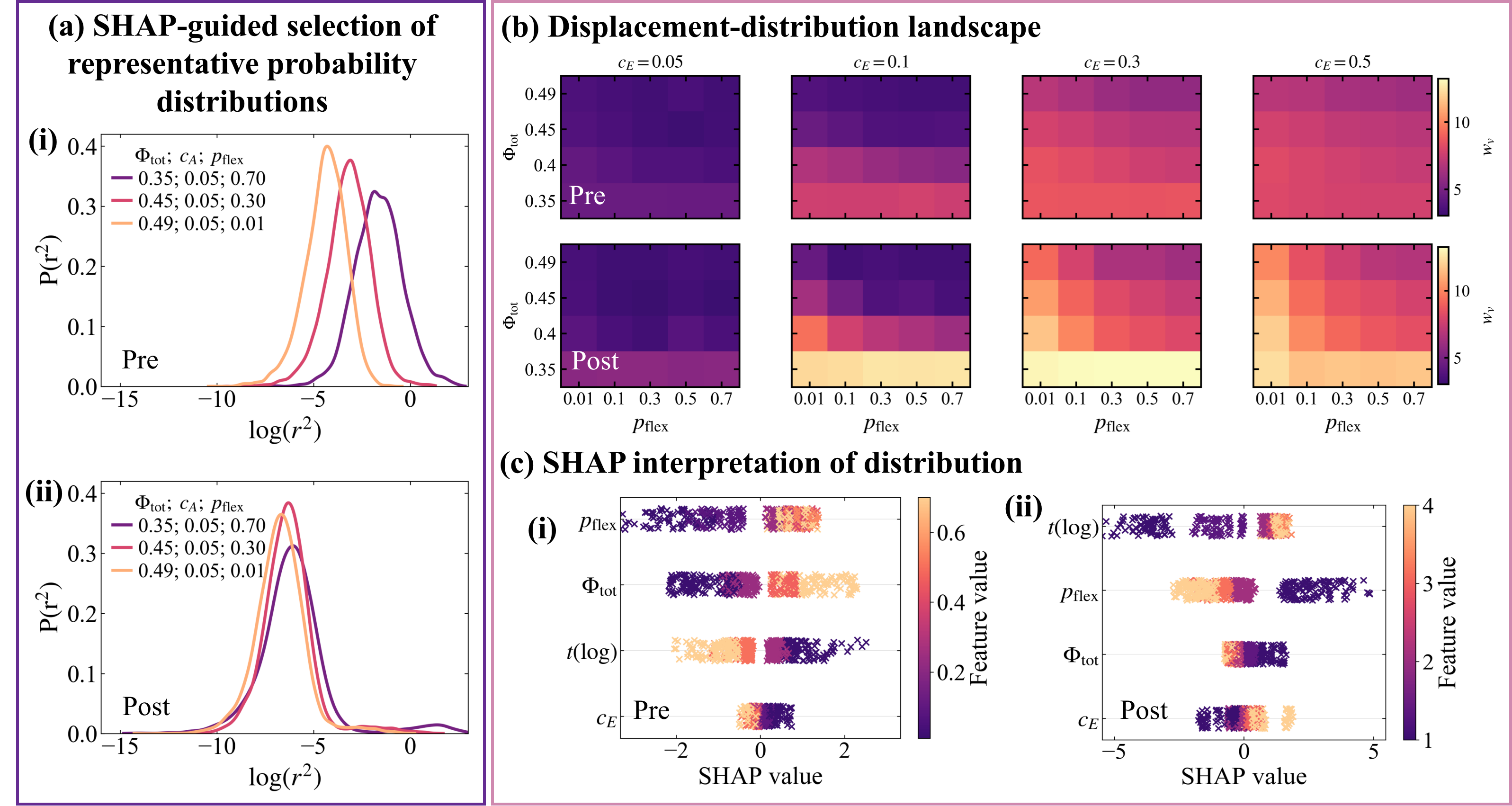}
	\caption{\textbf{SHAP-guided displacement distributions and their governing descriptors.} \textit{(a)} SHAP-guided selection of representative probability distributions of the explorers squared displacement, $P(r^{2})$, before (i) and after (ii) network formation for systems spanning low-, intermediate-, and high-confinement environments at a fixed explorer fraction ($c_E=0.05$), obtained through different combinations of the total volume fraction ($\Phi_{\mathrm{tot}}$) and bond flexibility ($p_{\mathrm{flex}}$). \textit{(b)} Displacement-distribution landscape pre- (top) and post- (bottom) template formation, showing the evolution of the distribution width, $w_v$, as a function of $\Phi_{\mathrm{tot}}$, $p_{\mathrm{flex}}$, and $c_E$. Each column corresponds to a different explorer fraction ($c_E=0.05$, $0.10$, $0.30$, and $0.50$), while the color scale represents the corresponding distribution width, $w_v$. \textit{(c)} SHAP interpretation of the displacement-distribution width pre- (i) and post- (ii) network formation. The beeswarm plots illustrate the contribution of each control parameter to $w_v$. Within each row, marker color encodes the corresponding descriptor, with darker and lighter shades indicating lower and higher values, respectively.}	
	\label{Figure:7}
\end{figure*}	

	To further characterize how template network formation modifies explorer dynamics beyond the average transport measures discussed above, we examine the probability distribution of the squared displacement, $P(r^2)$, which provides direct information on the distribution of particle mobilities within the system. Unlike the MSD and $F(q,t)$, which describe ensemble-averaged relaxation, the displacement distribution reveals how uniformly or heterogeneously explorers sample the available space under different confinement environments. Figure~\ref{Figure:7}a presents representative displacement distributions selected using the same SHAP-guided sampling strategy employed throughout this work, while maintaining a fixed explorer fraction ($c_E=0.05$). The selected systems span representative combinations of total volume fraction $\Phi_{\mathrm{tot}}$ and bond flexibility $p_{\mathrm{flex}}$, thereby allowing a direct comparison of how template network formation modifies the distribution of explorer displacements across distinct confinement environments. In the pre-network state, Fig.~\ref{Figure:7}a(i) shows that the three representative distributions remain single-peaked, but their maxima shift systematically toward smaller displacements as $\Phi_{\mathrm{tot}}$ increases and $p_{\mathrm{flex}}$ decreases. Thus, even before the formation of a percolated template network, the representative systems exhibit distinct displacement statistics, with the most probable explorer displacement depending sensitively on the underlying template environment. Following network formation (Fig.~\ref{Figure:7}a(ii)), the distributions remain single-peaked but become substantially more localized, with all three representative systems exhibiting pronounced peaks at smaller displacements than in their corresponding pre-network states. In contrast to the pre-network state, the representative distributions become more similar, indicating that network formation suppresses the large differences in the most probable explorer displacement observed during template formation. This systematic shift towards smaller displacements is consistent with the stronger caging inferred from the extended plateau in $F(q,t)$ and the reduced long-time MSD slopes, demonstrating that the establishment of a percolated template network restricts explorer motion across all representative confinement environments. 
	
	To quantify these trends across the complete parameter space, Fig.~\ref{Figure:7}b presents the displacement-distribution width, $w_v$, as a function of explorer fraction $c_E$ and bond flexibility $p_{\mathrm{flex}}$ for representative total volume fractions before and after template network formation. In all cases, $w_v$ increases systematically with increasing $c_E$, indicating that the displacement statistics become broader as the explorer population increases. This behavior is consistent with our previous study of irreversible binary colloids, where increasing $c_E$ led to the emergence of bimodal displacement distributions arising from the coexistence of slow and fast explorer populations~\cite{shireen2018cage}. In the present system, the representative distributions shown in Fig.~\ref{Figure:7}a remain predominantly single-peaked because they correspond to the low explorer fraction ($c_E=0.05$), while the global width analysis reveals that increasing $c_E$ continues to broaden the underlying displacement statistics throughout the parameter space. Conversely, increasing $\Phi_{\mathrm{tot}}$ systematically reduces $w_v$, reflecting the restricted range of explorer displacements in more crowded template environments. Following template network formation, the displacement-distribution width decreases systematically with increasing bond flexibility, indicating that flexible template networks produce a narrower distribution of explorer displacements than their more rigid counterparts. A reduction in the displacement-distribution width, $w_v$, indicates that explorer-particle displacements become confined to a narrower range, reflecting a more homogeneous displacement landscape. When considered together with the systematic shift of $P(r^2)$ towards smaller displacements (Fig.~\ref{Figure:7}a), the extended plateau in $F(q,t)$, and the reduced long-time MSD slopes, this narrowing provides complementary evidence for the enhanced localization of explorer particles following template network formation. Together, these results demonstrate that while bond flexibility modulates the extent of explorer localization within the percolated template network, the overall breadth of the displacement statistics remains primarily governed by the interplay between explorer concentration and crowding.

	Finally, the SHAP analysis in Fig.~\ref{Figure:7}c identifies the parameters governing the displacement-distribution width, $w_v$, before and after template network formation. Prior to network formation, bond flexibility and total volume fraction emerge as the dominant descriptors of $w_v$, while time and explorer fraction play comparatively smaller roles. This behavior reflects the fact that, during the aggregation stage, the evolving template morphology and the available free volume primarily determine the breadth of the explorer displacement distribution. Following network formation, the relative importance shifts markedly, with the observation time becoming the strongest predictor, followed by bond flexibility, indicating that the displacement statistics are increasingly governed by the long-time evolution of explorer dynamics within the established template network. The prominence of $p_{\mathrm{flex}}$ in both SHAP summaries indicates that bond flexibility contributes directly to the observed variation in the displacement-distribution width, although its role is stage dependent. Before network formation, this contribution reflects the evolving network environment, whereas after network formation it reflects the influence of local rearrangements within the percolated template network. Consequently, the SHAP analysis provides a quantitative link between the flexibility-dependent displacement statistics and the relaxation behavior established from the MSD and $F(q,t)$ analyses, complementing recent work relating evolving network morphology to dynamic heterogeneity through displacement-based analyses in soft polymer networks~\cite{mira2025characterizing}.	Collectively, the representative displacement distributions, the global width landscape, and the SHAP interpretation provide a unified picture in which template network formation systematically localizes explorer particles, while bond flexibility controls the extent of this localization across the explored parameter space.

 \section*{Summary and Conclusions}
 	
 	In this work, we investigated anomalous transport in irreversible binary colloidal systems with tunable bond flexibility by combining Brownian Cluster Dynamics simulations with a physics-guided interpretable machine-learning framework. Structural and dynamical behavior was systematically examined across a multidimensional parameter space defined by the total volume fraction, explorer fraction, and bond flexibility, pre- and post- template-network formation. The combined structural and dynamical analyses establish a consistent physical picture in which the evolution of the template network progressively strengthens explorer confinement, giving rise to increasingly sub-diffusive transport, delayed explorer-particle relaxation, and progressively localized displacement statistics. Collectively, the structural descriptors, mean-squared displacement, intermediate scattering function, and displacement distributions demonstrate how the coupled influence of crowding, composition, and bond flexibility governs the confinement environment and the resulting explorer-particle dynamics across the explored parameter space.
 	
 	A central physical finding of the present study is that bond flexibility acts as a dynamic regulator of explorer-particle relaxation within evolving colloidal networks. While crowding and composition determine the overall confinement environment, the influence of bond flexibility depends strongly on the stage of network evolution and the local template structure. Prior to template-network formation, relaxation is governed predominantly by crowding, with bond flexibility exerting only a comparatively weak influence. Following the formation of a percolated template network, however, bond flexibility becomes increasingly important by enabling local network rearrangements that facilitate the release of transient cages, particularly in explorer-rich systems. These results demonstrate that network flexibility provides an additional mechanism for regulating anomalous transport beyond the conventional controls of crowding and composition. More broadly, the present findings establish network mechanics as an important physical design variable for controlling transport and relaxation in heterogeneous soft-matter systems, with potential implications for colloidal gels, porous materials, biological transport, and other dynamically evolving disordered media.
 	
 	Beyond the physical insights into anomalous transport, the present work establishes a quantitative framework for investigating complex soft-matter systems through the integration of particle-resolved simulations with physics-guided interpretable machine learning. Rather than employing machine learning solely for prediction or post hoc interpretation, the framework is integrated directly into the scientific analysis to quantitatively disentangle the individual and coupled effects of the governing physical parameters across complementary structural and dynamical observables. This enables representative physical states to be identified objectively while providing a unified interpretation of structural evolution, transport, relaxation, and displacement statistics across the complete multidimensional parameter space. Consequently, physical relationships that were previously inferred qualitatively from selected simulation trajectories can now be quantified systematically across the entire parameter space. By transforming high-dimensional simulation data into a coherent mechanistic description of the underlying physics, the proposed integrated framework provides a quantitative methodology for investigating complex transport phenomena in computational soft-matter physics.
 	
	More broadly, the present work demonstrates how the integration of particle-resolved simulations with physics-guided interpretable machine learning can fundamentally change the way increasingly complex transport phenomena are investigated in soft-matter systems. Although demonstrated here using Brownian Cluster Dynamics simulations, the framework is readily extendable to other particle-resolved simulation methods and multidimensional experimental datasets in which multiple interacting physical variables govern transport behavior. By enabling quantitative attribution of coupled physical mechanisms while preserving direct physical interpretability, such integrated approaches provide a pathway toward simulation-experiment synergy, where interpretable machine learning not only assists physical understanding but also guides the systematic exploration of multidimensional parameter spaces. As simulation and experimental datasets continue to grow in complexity, physics-guided interpretable machine learning has the potential to become an important component of next-generation computational soft-matter research, accelerating mechanistic discovery while maintaining rigorous connections to the underlying physical principles.

	\section{Acknowledgment}
	We acknowledge Spartan HPC at The University of Melbourne, which provided the computing resources used in this work.
	
	\section*{Data availability}
	The data and analysis scripts supporting the findings of this study are available from the corresponding author upon reasonable request.
	
	\section{AUTHOR DECLARATIONS}
	\subsection{Conflict of Interest}
	The authors have no conflicts to disclose.
	
	\subsection{Author Contributions}
	Zakiya Shireen: Conceptualization; Investigation; Methodology; Writing-original draft; Writing-review and editing. 
	Sujin B Babu: Discussion, Writing-review and editing.

	\bibliography{reference}
\end{document}